\documentclass[preprint]{aastex}

\newcommand{\cc}{$\,\rm cm^{-3}$}

\newcommand{\cmsq}{$\,\rm cm^{-2}$}

\newcommand{\kmps}{km$\ \rm s^{-1}$}
\newcommand\HII{H\,{\sc ii}}
\newcommand\UCHII{UC\,H\,{\sc ii}}
\newcommand\vlsr{v$_{lsr}$}

\received{2003 October 25}
\begin{document}

\title{Methanol in W3(H$_2$O) and Surrounding Regions}

\author{E. C. Sutton\altaffilmark{1}, A. M. Sobolev\altaffilmark{2}, S. V. Salii\altaffilmark{2}, A. V. Malyshev\altaffilmark{2}, A. B. Ostrovskii\altaffilmark{2}, and I. I. Zinchenko\altaffilmark{3}}
\altaffiltext{1}{University of Illinois, 1002 W. Green St., Urbana, IL 61801; sutton@astro.uiuc.edu}
\altaffiltext{2}{Astronomical Observatory, Ural State University, Lenin Street 51, Ekaterinburg 620083, Russia; andrej.sobolev@usu.ru; svetlana.salii@usu.ru; malyshev@ampural.ru; osan@mail.ur.ru}
\altaffiltext{3}{Institute of Applied Physics of the Russian Academy of Sciences, Nizhny Novgorod 603950, Russia; zin@appl.sci-nnov.ru}

\begin{abstract}

We present the results of an interferometric study of
38 millimeter-wave lines of $^{12}$CH$_3$OH in the vicinity of
the massive star forming region W3(OH/H$_2$O).
These lines cover a wide range of excitation energies
and line strengths, allowing for a detailed study of
excitation mechanisms and opacities.
In this paper we concentrate on the region around the
water maser source W3(H$_2$O) and a region extending
about 30 arcsec to the south and west of
the hydroxyl maser source W3(OH).

The methanol emitting region around W3(H$_2$O)
has an extent of 2.0 x 1.2 arcsec (4400 x 2600 AU).
The density is of order 10$^7$ \cc, sufficient to
thermalize most of the methanol lines.
The kinetic temperature is approximately 140 K
and the methanol fractional abundance greater than 10$^{-6}$,
indicative of a high degree of grain mantle evaporation.
The W3(H$_2$O) source contains sub-structure,
with peaks corresponding to the TW source and Wyrowski's B/C,
separated by 2500 AU in projection.
The kinematics are consistent with these being distinct
protostellar cores in a wide binary orbit and a dynamical
mass for the region of a few tens of M$_{\sun}$.

The extended methanol emission to the southwest of W3(OH)
is seen strongly only from the lowest excitation lines and
from lines known elsewhere to be class I methanol masers,
namely the 84.5 GHz 5$_{-1}$--4$_0$ E
and 95.2 GHz 8$_0$--7$_1$ A$^+$ lines.
This suggests that this region, like class I maser sources,
is dominated by collisional excitation.
Within this region there are two compact clumps,
which we denote as swA and swB, each about 15 arcsec
(0.16 pc projected distance) away from W3(OH).
Excitation analysis of these clumps indicates the
presence of lines with inverted populations
but only weak amplification.
The sources swA and swB appear to have
kinetic temperatures of order 50--100 K
and densities of order 10$^5$--10$^6$ \cc.
The methanol fractional abundance for the warmer clump
is of order 10$^{-7}$, suggestive of partial grain mantle evaporation.
The clumping occurs on mass scales of order 1 M$_{\sun}$.

\end{abstract}

\keywords{ISM:clouds---ISM:individual (W3)---ISM:molecules---masers---radio lines:ISM}

\section{INTRODUCTION}

Methanol is an abundant interstellar molecule,
especially in regions of star formation where
its high abundance is thought to be the result of
thermal evaporation of dust grain mantles \citep{CHH93}.
A slightly asymmetric rotor with hindered internal
rotation and significant a- and b-axis dipole moments,
methanol has a complex spectrum which is sensitive to
interstellar conditions.  Observable lines cover
a wide range of energies and line strengths,
allowing for detailed analysis of excitation conditions
and opacities.

Two classes of interstellar methanol masers are known.
Class I methanol masers, produced by collisional pumping,
are seen in regions of massive star formation, but are
generally well separated from the centers of activity
as traced by embedded infrared sources and ultracompact
\HII\ (\UCHII) regions.
Class II methanol masers are thought to be produced
by radiative excitation at infrared wavelengths
and are spatially well correlated
with young stellar objects, \UCHII\ regions, and OH masers.
Even when not masing, methanol may be expected to exhibit
distinctly non-LTE excitation, except under the highest
density conditions where collisions could thermalize
the level populations.

The southeastern portion of W3 contains
the hydroxyl maser source W3(OH), which is associated
with an \UCHII\ region around a young O7 star,
and the nearby water maser source W3(H$_2$O), which contains
a young stellar object known as the TW object \citep{TW84}.
Numerous studies have been made of molecular material
associated with these two compact objects and
throughout the region surrounding them
\citep{W91,W93,W94,HvD97,W99,N00b}.
W3 is thought to be about 2.2 kpc from the Sun \citep{H78}.
At such a distance, resolution of order 1 arcsec is
necessary to study the nature of protostellar cores
on scales of order 1000 AU.

\cite{S01}\ presented results on new class II
methanol masers in W3(OH) and discussed excitation
conditions needed to explain observed fluxes in those
and several other methanol maser-candidate lines.
\cite{S03}\ extended that analysis
to include absorption and emission components in a larger set of
methanol lines in W3(OH) and discussed the implications
in terms of excitation and source structure.
Here we wish to complete our analysis of methanol in
the vicinity of W3(OH) and W3(H$_2$O) by presenting our
complete set of millimeter-wave observations and by
discussing the nature of the methanol associated with the
W3(H$_2$O) source and methanol in other nearby regions.

\section{OBSERVATIONS}

Observations were carried out with the BIMA\footnotemark\
interferometer between 1997 October and 2001 June,
following the general procedures discussed in \cite{S01}.
\footnotetext{The Berkeley-Illinois-Maryland Association operates
the BIMA array under support from the National Science Foundation.}
The set of observed $^{12}$CH$_3$OH transitions is presented in Table 1.
In addition, observations were made of the central four lines
of the J = 2--1 band of $^{13}$CH$_3$OH.
At 3 millimeters the data were primarily from the BIMA B and C arrays,
with a small amount of A array data for two of the lines.
At 1 millimeter the data were from the BIMA C and D arrays.
In all cases spatial resolution was of order 2 arcsec, as shown in Table 1.
Spectral resolution was generally of order 0.2 \kmps,
although some observations were made with resolution as wide as 1.2 \kmps.
The continuum was subtracted from the uv-data before maps were made.
The phase center of the maps is
$\alpha$(J2000) = 2$^\mathrm{h}$27$^\mathrm{m}$03\fs87,
$\delta$(J2000) = 61$\arcdeg$52$'$24\farcs6.
The flux density scale is thought to be accurate to within
about $\pm$ 10\% for the 3 mm band and about $\pm$ 25\% for the 1 mm band.

In order to check for spatially extended emission,
the 96.7 GHz methanol J = 2--1 band also was mapped using
the 20 meter Onsala telescope during 1998 January.
The BIMA J = 2--1 maps and the Onsala maps are in general agreement
with regards to the extent and shape of the emitting region.
Furthermore, the BIMA interferometer maps recover
about 75\% of the peak flux density of these lines in the Onsala data.
These lines are among the lowest energy lines in our data set,
and therefore more likely to be excited in diffuse,
spatially extended material.
Since we appear to have recovered most of the flux in these
low energy lines, we believe that we likely have done so
also for the remaining lines in our data set.
Since the BIMA maps have considerably higher spatial resolution
and positional accuracy and since we only have Onsala data
for four of our 38 lines,
in the discussion which follows we treat the BIMA maps by themselves,
without further reference to the Onsala data.

\section{RESULTS}

\subsection{Low energy lines}

In Figure 1 we show maps from those methanol lines
we have measured which have upper state energies
less than 100 K and for which we have u-v data which are
substantially complete down to spatial frequencies of 3 kilolambda or less.
In other words, these maps are based on data sufficient
to show any large scale structure which might be present.
Diffuse emission is present from spatially extended
material to the southwest of W3(OH).  This southwest region will be
discussed in detail in section 4.1, below.

\subsection{Other a-type lines}

In addition to the four J = 2--1 lines from
the ground torsional state, shown in Figure 1a,c,d,
we also observed all six lines from the J = 2--1
band in the first torsionally excited state
(v$_t$=1).  The torsionally excited
lines are shown in Figure 2.
These lines have upper state energies of order 300--400 K
and highlight the high excitation material in the
vicinity of the two main sources of luminosity,
W3(OH) and W3(H$_2$O).
Taken together, these ten a-type lines from the v$_t$=0,1
states form a standard against which other methanol lines
can be compared, since the a-type lines are less affected
by non-LTE conditions than the 28 b-type lines discussed below.

We also observed the
2$_0$--1$_0$ A$^+$, 2$_{-1}$--1$_{-1}$ E,
2$_0$--1$_0$ E, and 2$_1$--1$_1$ E lines of $^{13}$CH$_3$OH.
All four lines are weak, but visible towards
both W3(OH) and W3(H$_2$O).
Flux densities are consistent with $^{12}$C/$^{13}$C isotopic
ratios of order 60 and moderate opacities in the $^{12}$CH$_3$OH lines,
as discussed by A. M. Sobolev et al. (in preparation)
and in section 4.2, below.

\subsection{Other b-type lines}

In Figure 3 we present maps of 15 b-type methanol lines,
in addition to the 3 b-type lines already presented in Figure 1b,e,f.
Many of these lines are expected to be strong
under class II (radiative pumping) conditions.
These include the maser lines discussed in \cite{S01}\
(Fig. 3a,f,g) and the maser-candidate lines (Fig. 1b; Fig. 3c,e,h,i,l)
discussed in the same paper.  Note the tendency for the
emission to be stronger in these lines towards W3(OH)
than towards W3(H$_2$O), with the exceptions of the
6$_{-2}$--7$_{-1}$ E line at 85.5 GHz and the
8$_3$--9$_2$ E line at 94.5 GHz.
Both of these lines are predicted to mase under some
class II conditions by \cite{SCG97a},
and the 85.5 GHz line has been seen masing in G345.01+1.79
\citep{S98}.
However both lines were predicted to be in absorption under
the specific conditions of high kinetic temperature
derived by \cite{S01} for W3(OH).

Figure 3 also contains data on three new transitions
expected to be strong under radiative pumping conditions.
The 5$_1$--4$_2$ E line (Fig. 3b) has recently been seen
as a class II maser in G345.01+1.79 by \cite{S02},
and indeed it is quite strong here in the vicinity of W3(OH),
although there is no evidence of a narrow maser spike
at $-$43.1 \kmps, as was characteristic of the 86 and 107 GHz
masers discussed by \cite{S01}.
However, the W3(OH) emission extending from about $-$48 to
$-$43 \kmps\ could be a superposition of a number of weak maser features,
similar to the case of several maser-candidate lines
discussed in \cite{S01}.

The other new lines expected to be strong under
radiative pumping are torsionally excited lines,
1$_0$--2$_1$ E v$_t$=1 and 6$_1$--7$_2$ A$^-$ v$_t$=1.
These are particularly important in understanding the
class II maser excitation mechanism, since that mechanism
involves radiative excitation through the first two
torsionally excited bands of methanol, v$_t$=1 and 2.
\cite{V02} recently observed the 2$_0$--3$_1$ E v$_t$=1
line at 44.9 GHz in several galactic sources and concluded
that it is likely to be weakly masing in W3(OH).
They predicted that the closely related 1$_0$--2$_1$ E v$_t$=1 93.1 GHz line
would also be inverted in W3(OH), and our observations (Fig. 3m)
demonstrate strong emission in that direction.
Calculations show that 6$_1$--7$_2$ A$^-$ v$_t$=1 also
is highly excited by the class II mechanism and can
be a weak maser (A. M. Sobolev et al., in preparation).
We do not have enough evidence here to confirm masing
in either of these lines.  But the calculations mentioned above
indicate super-thermal excitation in both lines in the vicinity of W3(OH).

Other b-type lines are expected to be favored
under class I (collisional pumping) conditions.
The 5$_{-1}$--4$_0$ E and
8$_0$--7$_1$ A$^+$ lines will be discussed extensively in \S4.1.
Here in Figure 1e,f we draw attention to the fact that emission from
these lines is somewhat stronger from W3(H$_2$O) than from W3(OH).
Data are presented on three other lines from the
ground torsional state:
8$_0$--7$_1$ E,
11$_{-1}$--10$_{-2}$ E, and
13$_1$--12$_2$ A$^+$ (Figures 3d, j, and k, respectively).
It is unclear how strongly
8$_0$--7$_1$ E and 13$_1$--12$_2$ A$^+$
might be favored by class I conditions and disfavored by
class II conditions, although the data show that
13$_1$--12$_2$ A$^+$
emission is indeed stronger towards W3(H$_2$O) than W3(OH).
The 104.3 GHz 11$_{-1}$--10$_{-2}$ E line belongs to the same series
as the 9.9 GHz class I methanol maser 9$_{-1}$--8$_{-2}$ E.
To the best of our knowledge, the 9.9 GHz maser
has been seen, weakly, in only one source, W33-Met \citep{SKV93}.
Voronkov and Alakoz have recently seen maser emission in the 104.3 GHz line,
again only in W33-Met (M. Voronkov 2003, private communication).
Calculations reported by \cite{C01} suggest that the
relative excitation of the K = $-$1 and K = $-$2 E ladders
is very model dependent.  Here we simply note that
emission from 11$_{-1}$--10$_{-2}$ E is stronger
towards W3(OH) than W3(H$_2$O).
Finally in Figure 3n we present data from a
torsionally excited line, 6$_1$--5$_0$ E v$_t$=1, which
should be more favored by class I conditions than class II,
and indeed it is stronger towards W3(H$_2$O) than W3(OH).

We have data on ten other b-type lines, listed in Table 1
but not presented in Figures 1 or 3.
Although these additional lines will be used
in the analysis which follows, the maps either
have relatively poor sensitivity or the lines
are not strongly preferred by either class I or
class II conditions.  For example, four of these
lines are relatively high J members of the
J$_{-2}$--J$_1$ E species Q-branch.
Most of the remaining lines in this category are
of similarly high J or have $|$K$| \ge$ 3.
They are of minor importance in the analysis.

\section{DISCUSSION}

\subsection{Spatially extended material to the southwest}

Some emission to the southwest of W3(OH) is visible in all of
the lines shown in Figure 1, although it is most extensive in the lowest
energy lines plotted in Figure 1a.  This is a blend of the
2$_0$--1$_0$ A$^+$ (E$_\mathrm{u}$ = 7 K) and
2$_{-1}$--1$_{-1}$ E (E$_\mathrm{u}$ = 12.5 K) lines.
The relative weakness of the emission in this region
for the other lines of the J = 2--1 band,
2$_0$--1$_0$ E at E$_\mathrm{u}$ = 20 K (Fig. 1c) and
2$_1$--1$_1$ E at E$_\mathrm{u}$ = 28 K (Fig. 1d),
indicates this is optically thin emission from low excitation material.
An average of the emission in the lower right quadrant
of the plots for the J = 2--1 lines
(excluding emission within 4 arcsec of W3(OH) itself)
implies a rotational temperature of 11 K.
This is consistent with a kinetic temperature of order 50 K
and a density of order $10^5$--$10^6$ \cc\
(more detailed calculations are given below).
This methanol emission shows good spatial and kinematical correspondence
with material to the southwest of W3(OH) seen in
NH$_3$ (1,1) and (2,2) emission by \cite{W93},
who find the gas in that region to be characterized by
a hydrogen density of 6 $\times$ $10^5$ \cc\
and a rotational temperature of 30 K.
It appears that in methanol we are seeing material
similar to that seen in metastable ammonia.
The presence of high density material in this region
is also suggested by the SO and CS data of \cite{N00a}.

Within the low excitation methanol region are two clumps
which show up prominently in the
5$_{-1}$--4$_0$ E and 8$_0$--7$_1$ A$^+$ lines.
These are the clumps labelled swA and swB in Fig. 1e,f,
at offsets of ($\Delta\alpha=-8$, $\Delta\delta=-14$) and
($\Delta\alpha=-13$, $\Delta\delta=-7$) arcsec, respectively.
Since W3(OH) itself is a class II methanol maser source,
the presence of strong compact emission in these lines,
known for their strong \textbf{class I} maser emission, is unexpected.
It seems that in these two regions at least, the excitation
conditions are more like those found in class I sources.
The class I mechanism is driven by collisional pumping.
Therefore, in the regions of clumps swA and swB, collisions
evidently are producing super-thermal excitation in these, and
potentially other, class I methanol maser lines,
whereas radiative pumping is less important.
These positions are also near local maxima in
the 2$_0$--1$_0$ A$^+$ and
2$_{-1}$--1$_{-1}$ E map (Fig. 1a).
The class I mechanism can also modestly enhance the intensity
of these two particular J = 2--1 lines,
although it seems likely that these positions
are also local maxima in methanol column density.

To show that either of these regions is actually weakly masing
via the class I mechanism requires further evidence such as
sub-thermal linewidths or brightness temperatures well in
excess of conceivable kinetic temperatures.  Data on linewidths
and flux densities from clumps swA and swB are contained in Table 2.
The linewidths are narrow (2--3 \kmps\ FWHM for the $5_{-1}$--$4_0$ E
and $8_0$--$7_1$ A$^+$ lines), but not sufficiently so
to conclude that these are masers (linewidths of class II masers
in this source are typically 0.3 \kmps, and thermal widths
at 50 K are of similar order).

Emission is strong and compact for clump swA in the
8$_0$--7$_1$ A$^+$ line, so this would seem to be the best
case for making an argument in favor of maser emission
based on brightness.  The spatial resolution of the
map show in Figure 1f is 2.5 x 2.2 arcsec.
Clump swA is partially resolved in this map.
Based on a deconvolved source size of 3.0 x 1.9 arcsec
and a peak flux density of 2.0 Jy,
we derive a brightness temperature of 48 K.
If structure is present at spatial scales smaller than
the current resolution, this represents a lower limit on the brightness.
This is not sufficient to prove class I maser action,
but since this temperature is greater than that known for other
material to the southwest of W3(OH) and there is no known
source of luminosity near this position, it seems likely
that some degree of maser amplification is taking place.
The alternative, that this is simply a high density clump
in otherwise rather warm material, seems unlikely due to
the weakness of the 2$_0$--1$_0$ E and
2$_1$--1$_1$ E emission in this region.

The remaining line to be discussed in this context is
0$_0$--1$_{-1}$ E, presented as Figure 1b.
In this transition we see relatively little emission
at the locations of clumps swA and swB as well as throughout
the region to the southwest of W3(OH), even though
the upper state energy of this transition is only 13.1 K.
The reason for this appears to be that this line is
favored by class II excitation conditions and is a
known class II maser in G345.01+1.79 \citep{V99}.
Lines which mase or are super-thermally excited under
class II conditions are expected to be sub-thermally
excited under class I conditions such as prevail here
to the southwest of W3(OH) and specifically near clumps swA and swB.

It might seem difficult to be precise about the conditions
in these regions due to the small number of lines detected
towards clumps swA and swB, the likely non-thermal nature of the
excitation, and the possibility of significant opacity
in the lowest energy lines.
However, non-LTE fits to the data for these seven lines do provide
significant constraints on source parameters, as shown in Table 3.
These calculations were performed according to the methods
described by \cite{SCG97a}, \cite{S01}, and \cite{SSK02},
but without an incident infrared radiation field.
Assuming the warm dust is localized to the immediate vicinities
of W3(OH) and W3(H$_2$O), the infrared field should be down
by a factor of at least 500 at the positions of swA and swB,
relative to that near W3(OH).
The physical parameters of kinetic temperature, hydrogen density,
methanol column density, and source size were varied to obtain
optimal agreement between the model and the observed fluxes.
Parameter uncertainties were obtained in the usual fashion
\citep{P92} from the curvature of the $\chi^2$ hypersurface.
Each represents a 68\% confidence limit after marginalizing over
the remaining parameters.

Comparing our methanol column densities with an
estimated hydrogen column density of
$3 \times 10^{23}$ cm$^{-2}$ for this region \citep{W93},
we derive a methanol fractional abundance of order $10^{-8}$.
Independent estimates of hydrogen column densities for the
individual sources can be obtained by multiplying the hydrogen
number densities from Table 3 by the thicknesses along the
lines of sight (assumed to be equal to the derived source diameters
of 4400 AU for swA and 5000 AU for swB).
These yield methanol fractional abundances of
$3.7 \times 10^{-7}$ for swA and $3.1 \times 10^{-8}$ for swB.
These methanol fractional abundances are
higher than the values of order $10^{-9}$ usually seen
in dark molecular clouds \citep{KS94}.
But they appear to be lower than the values
of order $10^{-6}$ found for other star forming regions \citep{M86} and
generally attributed to thermal evaporation of dust grain mantles.
Thermal evaporation of methanol from polar ice mantles
is thought to occur at temperatures of 85 - 90 K
\citep{HH93,vD93,TW97,vDB98}.
Therefore the higher methanol fractional abundance in swA
is consistent with its higher temperature, as shown in Table 3.
At the minimum projected distances of these clumps from W3(OH),
neither gas nor dust grains are likely to reach
temperatures of this magnitude due solely to the
luminosity of W3(OH) \citep{DN97}.
Therefore it is possible that swA contains an internal heating source.
Since the temperature of swB is insufficient to evaporate methanol,
some non-thermal desorption mechanism may be necessary to release
the observed amount of methanol.

Assuming uniform densities, masses for the gas in clumps
swA and swB also may be obtained, as shown in Table 3.
These masses are of order 0.05 M$_{\sun}$ and 1.1 M$_{\sun}$, respectively,
with large uncertainties which are probably dominated by uncertainties
in the sizes and filling factors.
These numbers may be compared with an estimate for the total mass
in a larger region to the southwest of W3(OH) of
80 M$_{\sun}$ \citep{W93}.
The observed linewidths imply virial masses of order
25 M$_{\sun}$ each for swA and swB.
It is unlikely that either clump is gravitationally bound,
unless it contains a compact core of significant mass
which remains unresolved in the current observations.
A collapsing high mass core or a high mass protostar
at a very early evolutionary stage, in the center of swA,
could provide sufficient mass to bind the surrounding gas cloud
as well as the luminosity needed to produce the
observed gas temperature.

For source swA, our best model fit shows that all of the lines
listed in Table 2 are inverted, except for $0_0$--$1_{-1}$ E.
However, only $5_{-1}$--$4_0$ E and $8_0$--$7_1$ A$^+$
show significant amplification, with $\tau = -0.8$
and $\tau = -0.9$, respectively.
For source swB the energy levels are more nearly thermalized,
as expected for the higher H$_2$ density.  Although three of
the lines are still inverted, only $8_0$--$7_1$ A$^+$
shows significant amplification, with $\tau = -0.6$.
Our model predicts that at these locations the closely related
36 GHz ($4_{-1}$--$3_0$ E) and 44 GHz ($7_0$--$6_1$ A$^+$)
methanol maser lines will also be inverted
and show modest amplification ($\tau \approx -1$).
Higher J members of these series exhibit weaker inversions.
Observed line widths of 2.1 \kmps\ \citep[36 GHz]{HB89},
0.8 \kmps\ \citep[44 GHz]{HMB90}, 2.1 \kmps\ (for swA at 84.5 GHz; this work),
and 2.5 \kmps\ (for swA at 95.2 GHz; this work) are narrower than
typical widths for this source \citep{K97},
consistent with modest amplification in these lines.

Our modelling of the southwest region gives a possible explanation
for some of the complex behavior of the 25 GHz J$_2$--J$_1$ E lines
near W3(OH).  The observations of \cite{M86} showed a mixture of
absorption and emission line components in this band
when viewed with a large beam towards W3(OH).
The strongest emission was seen for J=3,
although emission was evident up to J=9.
Our model of swB predicts a weak inversion in the J=3 line,
super-thermal excitation in J=4, and
weak emission for other nearby values of J.
In contrast, \cite{S01} predicted absorption from
the material directly in front of the W3(OH) \UCHII\ region.
Since the \cite{M86} beam encompassed both W3(OH) and swB,
their observations showed a mixture of these characteristics,
further complicated by emission and absorption from other
methanol-rich regions contained within their beam (cf. Fig. 1a).

\subsection{W3(H$_2$O) region}

Most of the lines studied here exhibit strong emission
from the region around the water maser source W3(H$_2$O).
Careful examination of our maps with the best signal-to-noise
ratio and highest spatial resolution reveal that the
bulk of the methanol emission comes from a region approximately
2.0 x 1.2 arcsec in size, elongated along a position angle
of $100\arcdeg \pm 10\arcdeg$ and centered about 5.3 arcsec east of our
nominal map center
($\Delta \alpha = 5\farcs3 \pm 0\farcs1 , \Delta \delta = 0\farcs1 \pm 0\farcs1$).
In other words, the emission comes
from a region encompassing the TW object \citep{TW84}\
and continuum peaks B \& C of \cite{W99}.

This is illustrated in Figure 4, which presents
a 0.6 arcsec resolution map of the 1.3 mm continuum from this region.
Sources A (the TW object), B, and C of \cite{W99}\ are labelled.
Our continuum map has slightly lower resolution than Wyrowski et al.,
and we do not resolve their sources B and C from each other.
A solid line is overlaid on the map to represent the
elliptical region of methanol emission described above.
The extent of the methanol emission roughly corresponds to
the 0.2 Jy beam$^{-1}$ contour of the continuum map.
The similarity in position, size, shape, and orientation
of the methanol and the continuum emission strongly
suggests they are coming from the same material.
We note that the agreement in position can be made even better
by shifting the 1.3 mm continuum map south 0.1 arcsec and west 0.1 arcsec.
Such a shift is within the estimated positioning error of our map
and also serves to shift the center of the W3(OH) 1 mm continuum
onto our previously reported 3 mm position, as shown \citep{S01}.
It also brings our two W3(H$_2$O) peaks into better agreement with
peak A and an average of peaks B and C of \cite{W99}.

In Table 4 we present the observed integrated line fluxes
for these lines, and in Figure 5a we plot these fluxes
versus upper state energies in the usual "rotation diagram" format.
The diagram assumes a source solid angle of
$\Omega = 4.4 \times 10^{-11}$ ster, a value which gives
the optimal fit in the procedure described below.
In the limit of optically thin emission from a region in LTE,
the points in a rotation diagram should fall on a straight line
with the slope indicating the kinetic temperature and the intercept
containing information about the total column density, according
to the relations
\[
\frac{\mathrm{N}_\mathrm{u}}{\mathrm{g}_\mathrm{u}} =
\frac{\mathrm{3c}^2}{16\pi^3\nu^3} \; \frac{1}{\Omega} \;
\frac{\int \mathrm{S}_{\nu} \mathrm{dv}}{\mu^2 \mathrm{S}}
\]
and
\[
\frac{\mathrm{N}_\mathrm{u}}{\mathrm{g}_\mathrm{u}} =
\frac{\mathrm{N}_\mathrm{T}}{\mathrm{Q(T}_{\mathrm{rot}})} \;
e^{- \mathrm{E}_\mathrm{u}/\mathrm{kT}_{\mathrm{rot}}} \; .
\]
In these equations the total column density, N$_\mathrm{T}$, and
the upper state column density divided by the level degeneracy,
N$_\mathrm{u}$/g$_\mathrm{u}$, are averages
over the source solid angle, $\Omega $.
The rotational partition function is Q(T$_{\mathrm{rot}}$), and
the vibrational partition function is neglected.
The observed flux integrated over velocity is $\int \mathrm{S}_{\nu}$dv.
The energy of the upper state is E$_\mathrm{u}$,
and the square of the dipole moment matrix element
for the transition is $\mu^2$S.
Values for both are given in Table 1.
The other symbols take on their usual meanings.
The data show large deviations from straight line behavior,
indicating the breakdown of one or more of the assumptions
implicit in the rotation diagram technique.
Several of the lines shown are in fact optically thick.
In a rotation diagram, optically thick lines appear
artificially low since the observed intensity is less than
that expected from a linear extrapolation from the optically
thin regime.  The optical depth of a line in LTE is given by
\[
\tau = \frac{8\pi^3}{\mathrm{3h}} \; \mu^2\mathrm{S} \;
\frac{\mathrm{N}_0}{\mathrm{g}_0} \;
e^{- \mathrm{E}_\mathrm{u}/\mathrm{kT}_{\mathrm{rot}}} \;
(e^{\mathrm{h}\nu/\mathrm{kT}_{\mathrm{rot}}}-1) \;
\frac{1}{\Delta \mathrm{v}} \; ,
\]
where $\Delta $v is the linewidth and N$_0$/g$_0$
is the column density of the ground state.  The optical depths of these
lines range from 0.1 to 4.6, with approximately half of the lines
showing significant saturation.  The effects of saturation may
be corrected using the factor C$_{\tau}$ \citep{GL99,G00}, where
\[
\mathrm{C}_{\tau} = \frac{\tau}{1-e^{-\tau}} \; .
\]
In Figure 5b we present a "corrected" rotation diagram,
in which the deviations from straight line behavior are much reduced.
It is apparent that the largest deviations in the original diagram
were the result of saturation.
The best fit in Figure 5b has an intercept of
N$_0$/g$_0$ = 1.3 $\times$ $10^{15}$ \cmsq,
a slope corresponding to a rotational temperature of 149 K,
and a solid angle of $4.4 \times 10^{-11}$ ster.
This gives a value of $\chi^2$ = 128 for 35 degrees of freedom
(38 data points minus three free parameters).
Some significant deviations from a good fit still remain,
so the next step is to determine whether they represent
real departures from LTE.

A non-LTE environment in which collisional pumping dominates over
radiative excitation produces conditions similar to those
in class I methanol masers.  Under these conditions, certain
lines are predicted to have super-thermal excitation, particularly
those with upper states in the K = $-$1 E and K = 0 A ladders.
Four out of the five such lines in this data set lie above
the average trend of the data in Figure 5b.  Similarly, b-type lines with
lower states in these ladders are predicted to show sub-thermal
excitation.  Again four out of five such lines lie below
the average trend of the data.  Therefore, the conditions
around W3(H$_2$O) appear similar to those in class I maser regions,
even though W3(H$_2$O) is not a methanol maser source.

We performed detailed non-LTE calculations for the data in Table 4,
using a simplified model of the infrared radiation field.
Such a treatment is justified since at these densities
methanol is near the collisionally dominated limit.
\cite{OS02} have shown that when the color temperature of
the radiation is in the range 125--250 K, the methanol radiative excitation
occurs primarily through a large number of strong methanol
absorption lines in the 15--40 $\mu$m wavelength range.
The most important parameter of the infrared radiation field
therefore is its intensity in this band.
We assume that the pumping radiation is produced by warm dust
intermixed with the gas and at the same physical temperature as the gas.
At these wavelengths the dust is optically thick in this source,
minimizing the importance of infrared sources external to the cloud
(although the TW source and Wyrowski's B/C remain the ultimate
sources of energy in this region).

Methanol radiative transition rates were taken from \cite{MGH99}.
Collisional transition rates were based on the model of \cite{PW93}.
For simplicity, a $\lambda^{-2}$ dust opacity law was chosen.
Details of the excitation do depend on the optical properties
of the dust grains \citep{OS02},
but the differences are small at high gas densities.
Radiative transfer for the millimeter-wavelength methanol lines
was treated using the large velocity gradient (LVG) approximation.
Further details regarding the calculational methods may be
found in \cite{KK80}, \cite{S82}, \cite{SCG97a}, and \cite{SCG97b}.

The gas cloud was assumed to be uniform in H$_2$ density,
gas and dust temperature, dust-to-gas ratio, and
methanol fractional abundance.
The data were not adequate to justify a more complicated,
spatially inhomogeneous model of the cloud.
The source solid angle was fixed at $4.4 \times 10^{-11}$ ster, as given above.
The other four parameters were allowed to vary.
An optimal fit for W3(H$_2$O) was obtained with
a kinetic temperature of $140 \pm 10$ K,
hydrogen density of $0.9^{+2.4}_{-0.7} \times 10^7$ \cc,
and N$_{\rm A}$ + N$_{\rm E}$ = $2.4^{+0.3}_{-0.3} \times 10^{18}$ cm$^{-2}$,
as shown in Table 3.
We can multiply the hydrogen number density
by the estimated source thickness of 3400 AU
to derive a hydrogen column density.
Combining this with the measured methanol column density gives
a methanol fractional abundance of order $5 \times 10^{-6}$ for W3(H$_2$O).
In comparison with the values in Table 3, \cite{vdT00a}
give a kinetic temperature of greater than 200 K for W3(H$_2$O)
from statistical equilibrium calculations for H$_2$CO,
similar to those given earlier by \cite{H94} and \cite{HvD97}.
However, the beam size for those observations was large enough
to encompass W3(OH) as well as W3(H$_2$O).
The methanol column density of \cite{vdT00b} is smaller than
that given in Table 3, an effect we also attribute to
their larger beam size.
The density profile of \cite{vdT00a} corresponds to a density of
about $3 \times 10^7$ cm$^{-3}$ at a characteristic radius of 2000 AU,
consistent with our result.

The observations also provided good constraints on the amount of
dust needed in the model.  This gives us an independent method
to calculate the hydrogen column density and derive a methanol
fractional abundance.
Previously we adopted a $\lambda^{-2}$ dust opacity law.
Now, only for the purpose of this calculation, we further
adopt a gas-to-dust mass ratio of 100 and
a dust mass absorption coefficient of 1.1 cm$^2$ g$^{-1}$
at 1.3 mm \citep{H95}.
This yields a methanol fractional abundance of order
$1.7 \times 10^{-6}$ for W3(H$_2$O).
Given the uncertainties inherent in these various assumptions,
the two estimates of the fractional abundance agree well
and support the notion of significant grain evaporation in this source.

Figure 5c shows the residuals from the optimal non-LTE fit.
For this fit $\chi^2$ = 86 for 34 degrees of freedom
(38 data points minus 4 free parameters).
The non-LTE model is a significantly better fit to the data
than the LTE model.  There are still some significant residuals,
possibly indicating that the observational uncertainties were underestimated.
But it should be kept in mind that there are additional uncertainties,
those associated with the model calculations.
Such uncertainties are difficult to quantify,
but they appear to be of the same order as those associated
with the observations.  The nature of these uncertainties
is discussed in Appendix B of \cite{S01}.
The dominant uncertainty may be spatial variation in the source parameters.

The results presented above are based on averages over
the entire W3(H$_2$O) source.  Although our spatial resolution
is generally insufficient to resolve the TW object from
sources B and C of \cite{W99}, we do have some
information on spatial sub-structure from the shape of the spectra.
For the 29 v$_t$=0 lines listed in Table 1,
the average methanol emission velocity towards W3(H$_2$O)
is $-$50.0 \kmps, with a standard deviation of 0.37 \kmps\
(giving a standard error of the mean of less than 0.1 \kmps).
Several lines with good signal-to-noise ratios
show distinct shifts from this velocity, as shown in Figure 6.
The 6$_1$--7$_2$ A$^-$ v$_t$=1 line is well fit
by a single gaussian with a velocity of $-$51.0$\pm$0.1 \kmps.
This blueshifted emission is more characteristic of
the TW source in the eastern part of W3(H$_2$O).
Note that in Figure 3o the emission peak for this line
is spatially coincident with the TW object.
This is a high excitation line (E$_\mathrm{u}$ = 374 K),
presumably more easily excited close to the main
source of luminosity in the region, namely the TW object.
The 8$_0$--7$_1$ A$^+$ and 5$_{-1}$--4$_0$ E lines
have mean velocities of $-$49.3$\pm$0.1
and $-$49.5$\pm$0.1 \kmps, respectively.
Redshifted emission is characteristic of the western portion
of W3(H$_2$O), near Wyrowski's source C.
And in the maps of these lines, the emission peaks distinctly
to the west of TW (Figures 1e,f -- although this shift to the west
is too small to be seen on the scale at which these maps are presented).
So the data suggest that the line-of-sight velocity
difference between the TW object and source C may be
of order 1.6 \kmps.
The 8$_0$--7$_1$ A$^+$ and 5$_{-1}$--4$_0$ E lines
are the same two lines discussed in section 4.1,
known elsewhere for their class I methanol maser emission.
Since such lines are known to be collisionally pumped,
the luminosity source proposed by \cite{W99} may either
be sufficiently weak or sufficiently well shielded
from the methanol-rich material for collisional pumping
to dominate.  We should note that in our non-LTE model
for the W3(H$_2$O) region, these two lines are optically thick,
with opacities of $\sim$3.7 and $\sim$1.6, respectively.
So we may be seeing a preponderance of foreground material
in these lines.

Given a density of $0.9 \times 10^7$ \cc\ and assuming
that W3(H$_2$O) is as deep along the line of sight
as its average extent in the plane of the sky (3400 AU),
this corresponds to 1.6 M$_{\sun}$ of warm gas,
excluding that contained in proto-stellar sized
objects and condensed cores.
This may be compared with a previous estimate of the gas mass
of "at least a few M$_{\sun}$" \citep{TW84}.
The estimate of 70 M$_{\sun}$ by \cite{W91} refers to
a considerably larger region (0.13 pc) probed by C$^{18}$O.
\cite{W94} estimate 60 M$_{\sun}$ in a 0.05 pc diameter region
from C$^{18}$O observations; however,
they also estimate 30 M$_{\sun}$ on a 0.01 pc scale
from dust observations.
Finally, \cite{W93} give a dynamical mass of 18 M$_{\sun}$,
which therefore includes that of any protostars.

Non-thermal continuum emission at centimeter and decimeter wavelengths
has been seen both east and west of the TW object \citep{R95,WRM99,ARM03}.
This has been interpreted as synchrotron radiation
from precessing jets \citep{WRM99}, in part due to the
symmetry and curvature of the emission and its location
near the apparent center of expansion of the H$_2$O masers.
However, alternative explanations are possible.
\cite{TPGA02} point out that in some cases of star formation,
water masers are found in protoplanetary accretion disks,
oriented perpendicular to the radio jet directions
(although \cite{TPGA02} do not consider the TW object
to be such a system).  But recently \cite{SS03} have
proposed a model in which the H$_2$O masers in W3(H$_2$O)
are formed on the surface of an east-west circumstellar disk
or ring.

This question of orientation has some bearing on the nature
of the object we have been calling Wyrowski's source C.
Its location is along the east-west axis of of
the synchrotron jet proposed by \cite{R95}.
In fact, it is roughly coincident with the cluster of
water masers located 1 arcsec west of the TW object.
Since jets produce shocks where they run into ambient material,
and shocks, in turn, can power masers,
one possibility is that Wyrowski's source C is a deeply
embedded Herbig-Haro object.
However, \cite{WRM99} and \cite{W99} agree that
regardless of jet geometry, the  most likely interpretation is
that Wyrowski's source C is a deeply embedded protostellar object.
We concur.

Using this interpretation and assuming that the molecular gas
seen in methanol is roughly equally divided between the TW object
and Wyrowski's source C, we get the following picture.
The TW object itself is a protostar of about 15 M$_{\sun}$ \citep{MWH88}.
Wyrowski's sources B/C may contain additional protostellar cores
of similar or smaller mass \citep{W99}.
Within 0.01 pc of each of these sources there is about
one solar mass of warm gas.
And on larger scales there are an additional few tens
of solar masses of cooler, less dense gas.

The dynamical stability of such a system is uncertain.
The putative sum of the protostellar masses (20-30 M$_{\sun}$),
the projected separation ($\approx$2500 AU),
and the estimated line of sight velocity difference (1.6 \kmps)
are consistent with a reasonable binary orbit
of the TW object with Wyrowski's B/C.
For example, a low eccentricity, high inclination,
intermediate phase (an angle of order 45\degr\
with respect to the line of nodes)
orbit is consistent with the above values.
This assumes that the 1.6 \kmps\ velocity difference
discussed above in connection with Figure 6
represents the true difference in line of sight protostellar velocities.
This would be the case if each protostellar object
were accompanied by its own circumstellar material,
but not necessarily the case if they were orbiting
within a common, circumbinary envelope.
A numerical simulation by \cite{TCBDPW95}\ produces a binary system
with circumstellar material in the form of
individual protostellar disks, a possible case of
the system described above.
\cite{LO2000} indicate that individual disks of sufficient size,
like those of \cite{TCBDPW95}, can be stable in binary systems.
The full width to zero power linewidths seen here are 10-12 \kmps.
Such velocities could be produced by slow infalls or outflows
or by orbital motion in individual circumstellar envelopes.
A significant amount of circumbinary material is also possible,
although if it is in purely orbital motion it could not account
for the full widths of these lines.

In a single dish study of high mass star forming regions,
\cite{MB02} report emission from the vicinity of W3(OH/H$_2$O)
in a number of methanol lines.  They report emission peaks
at $-51$ \kmps\ in the 6$_{-2}$--7$_{-1}$ E (85.5 GHz)
and 7$_2$--8$_1$ A$^+$ (111.2 GHz) lines.
Our work shows that gas at this velocity is clearly
associated with the W3(H$_2$O) source.
\cite{MB02} also point out that W3(OH/H$_2$O) is unusual among
massive star forming regions in not displaying clear maser
emission in the 8$_0$--7$_1$ A$^+$ (95.1 GHz) line.
The emission in this line from sources swA and swB was discussed
above in section 4.1 and that from W3(OH) by \cite{S03}.
For the W3(H$_2$O) source our model suggests an excitation temperature
for this line of about 135 K and an optical depth
of about 3.7, consistent with the lack of maser emission.

In our optimal non-LTE model of W3(H$_2$O) none of the lines
studied here are inverted; excitation temperatures range from
about 90--530 K.  Optical depths of these lines range
from about 0.07 to 3.9.

In our optimal non-LTE model, the lines in the J = 2--1
band of $^{12}$CH$_3$OH have opacities ranging from 1 to 1.5.
This means that, even apart from non-LTE effects,
the flux ratios for the J = 2--1 band of $^{12}$CH$_3$OH
relative to $^{13}$CH$_3$OH would be expected to be smaller than
the actual abundance ratio by a factor of 1.9.
In addition, since the $^{13}$CH$_3$OH lines are optically thin,
$^{13}$CH$_3$OH will be more efficiently cooled,
resulting in higher fractional populations in the lower rotational levels.
A detailed calculation of this effect would require
a larger set of $^{13}$CH$_3$OH observations and a separate
non-LTE model for $^{13}$CH$_3$OH.
However, the cooling is likely to be substantial.
The [$^{12}$C]/[$^{13}$C] abundance ratio is thought to be
$66 \pm 4$ \citep{LP90}.
Due to the above effects, the integrated flux ratio
for the J = 2--1 band should be considerably smaller,
possibly consistent with the observed ratio of $12 \pm 3$,
ignoring fractionation.

\section{Conclusions}

To the southwest of W3(OH) there is an extensive region of gas,
dense enough to be seen in a number of methanol lines.
Emission is most readily seen in lines of low upper state energy
(E$_\mathrm{u}$ $\lesssim $ 15 K) and in lines known elsewhere
for their class I methanol maser emission.
In this case, strong class I masers do not seem to be present,
but the physical conditions are similar to those seen
in class I maser regions.  Excitation is collisionally driven,
with densities of order 10$^5$ - 10$^6$ \cc.
Inverted populations and weak amplification appear to be present.
We predict that the 36 GHz and 44 GHz lines will also be inverted,
although it is doubtful that their optical depths will be
sufficient to produce strong masers.
The region is clumpy on scales of about 1 to a few tens of solar masses.
Kinetic temperatures range up to about 100 K.
The warmest regions have methanol fractional abundances
of order 10$^{-7}$, suggesting that some grain mantle evaporation
has taken place, but that the evaporation is incomplete.

In contrast, towards W3(H$_2$O) the methanol emission
is nearly thermalized and many lines are optically thick.
None of the lines studied have inverted populations.
Excitation is due to a combination of collisional
and radiative processes.
The size of the methanol emitting region is about 2600 x 4400 AU.
This is resolved into at least two components, corresponding
to the TW object and a blend of Wyrowski's B and C components.
It is likely that both are massive young stellar objects.
Their separation and line of sight velocity difference are
consistent with a wide binary orbit with a period of order
50,000 years.  Each source may be accompanied by circumstellar
material containing of order one solar mass or less of gas.
There may be as much as a few tens of M$_{\sun}$ of
circumbinary material, although its distribution is unclear.
Methanol fractional abundance in the circumstellar material
is of order 10$^{-6}$, indicative of a warm environment and
extensive grain mantle evaporation.

\acknowledgments
This work was supported in part by the National Science Foundation
under BIMA grant AST 99-81363 and CARMA grant AST 02-28953.
A. M. S., S. V. S., A. V. M., A. B. O., and I. I. Z.
were supported by INTAS grant 99-1667, RFBR grant 03-02-16433,
and the Russian Ministry of Education (grant E02-11.0-43).
The authors thank Yuri Shchekinov for helpful discussions.

\clearpage


\begin{deluxetable}{lcllc}
\tabletypesize{\footnotesize} \tablewidth{245pt} \tablecolumns{6}
\tablenum{1} \tablecaption{Observed $^{12}$CH$_3$OH Lines}
\tablehead{
\colhead{Transition} & \colhead{Frequency} & \colhead{E$_u$} & \colhead{$\mu^2$S} & \colhead{Beam} \\
& \colhead{(MHz)} & \colhead{(K)} & \colhead{(D$^2$)} &
\colhead{(arcsec$^2$)} } \startdata
7$_2$--8$_1$ A$^-$   & $\;\;$80993.26   & 103. & 2.52 & 2.6 x 2.5\\
13$_{-3}$--14$_{-2}$ E & $\;\;$84423.71 & 274. & 4.30 & 2.5 x 2.4\\
5$_{-1}$--4$_0$ E    & $\;\;$84521.21   & $\;\;$40. & 3.08 & 2.5 x 2.4\\
6$_{-2}$--7$_{-1}$ E & $\;\;$85568.07   & $\;\;$75. & 2.01 & 3.7 x 2.7\\
7$_2$--6$_3$ A$^-$   & $\;\;$86615.60   & 103. & 1.36 & 2.0 x 1.7\\
7$_2$--6$_3$ A$^+$   & $\;\;$86902.95   & 103. & 1.36 & 2.0 x 1.8\\
8$_{-4}$--9$_{-3}$ E & $\;\;$89505.78   & 171. & 1.56 & 3.6 x 2.6\\
1$_0$--2$_1$ E $\phn\phn$ v$_t$=1 & $\;\;$93196.66 & 303. & 1.34 & 2.3 x 2.1\\
8$_3$--9$_2$ E       & $\;\;$94541.81   & 131. & 2.24 & 2.8 x 2.2\\
8$_0$--7$_1$ A$^+$   & $\;\;$95169.52   & $\;\;$84. & 7.22 & 2.5 x 2.2\\
2$_1$--1$_1$ A$^+$ $\phd$ v$_t$=1 & $\;\;$96396.06 & 332. & 1.21 & 2.1 x 1.9\\
2$_1$--1$_1$ E $\phn\phn$ v$_t$=1 & $\;\;$96492.16 & 298. & 1.21 & 2.1 x 1.9\\
2$_0$--1$_0$ E $\phn\phn$ v$_t$=1 & $\;\;$96493.55 & 308. & 1.62 & 2.1 x 1.9\\
2$_{-1}$--1$_{-1}$ E  v$_t$=1 & $\;\;$96501.70 & 420. & 1.21 & 2.1 x 1.9\\
2$_0$--1$_0$ A$^+$ $\phd$ v$_t$=1 & $\;\;$96513.67 & 431. & 1.62 & 2.1 x 1.9\\
2$_1$--1$_1$ A$^-$ $\phd$ v$_t$=1 & $\;\;$96588.59 & 332. & 1.21 & 2.3 x 2.0\\
2$_{-1}$--1$_{-1}$ E & $\;\;$96739.36   & $\;\;$12.5 & 1.21 & 2.1 x 1.8\\
2$_0$--1$_0$ A$^+$   & $\;\;$96741.38   & $\;\;\;\;$7.0 & 1.62 & 2.1 x 1.8\\
2$_0$--1$_0$ E       & $\;\;$96744.55   &  $\;\;$20. & 1.62 & 2.1 x 1.8\\
2$_1$--1$_1$ E       & $\;\;$96755.51   &  $\;\;$28. & 1.24 & 2.1 x 1.8\\
6$_1$--5$_0$ E $\phn\phn$ v$_t$=1 & $\;\;$99730.96 & 340. & 3.14 & 2.1 x 1.8\\
13$_2$--12$_3$ E     & 100638.87        & 234. & 3.84 & 3.2 x 2.4\\
13$_{-3}$--12$_{-4}$ E & 104060.72      & 274. & 3.27 & 3.7 x 2.5\\
11$_{-1}$--10$_{-2}$ E & 104300.40      & 159. & 3.41 & 3.1 x 2.3\\
13$_{-2}$--13$_1$ E  & 104336.64        & 237. & 1.20 & 2.9 x 1.5\\
10$_4$--11$_3$ A$^-$ & 104354.86        & 208. & 2.48 & 3.1 x 2.3\\
10$_4$--11$_3$ A$^+$ & 104410.49        & 208. & 2.48 & 3.1 x 2.3\\
13$_1$--12$_2$ A$^+$ & 105063.76        & 224. & 4.27 & 2.3 x 1.9\\
14$_{-2}$--14$_1$ E  & 105576.39        & 270. & 1.79 & 2.3 x 1.9\\
3$_1$--4$_0$ A$^+$   & 107013.81   & $\;\;$28. & 3.01 & 1.6 x 1.4\\
15$_{-2}$--15$_1$ E  & 107159.92        & 305. & 2.61 & 1.8 x 1.6\\
0$_0$--1$_{-1}$ E    & 108893.93        & $\;\;$13.1 & 0.98 & 2.2 x 1.8\\
14$_5$--15$_4$ E     & 109138.69        & 380. & 3.41 & 2.2 x 1.9\\
16$_{-2}$--16$_1$ E  & 109153.21        & 342. & 3.68 & 2.2 x 1.8\\
7$_2$--8$_1$ A$^+$   & 111289.60        & 103. & 2.34 & 1.9 x 1.6\\
5$_1$--4$_2$ E       & 216945.56        & $\;\;$56. & 1.12 & 2.5 x 2.3\\
6$_1$--7$_2$ A$^-$ $\phd$ v$_t$=1 & 217299.16  & 374. & 4.66 & 2.5 x 2.3\\
8$_0$--7$_1$ E       & 220078.49        & $\;\;$97. & 3.45 & 2.4 x 2.2\\
\enddata
\tablerefs{Xu and Lovas 1997, except the 107103.81 MHz line is
from Tsunekawa et al. 1995.}
\end{deluxetable}

\clearpage


\begin{deluxetable}{lccccc}
\tabletypesize{\small} \tablewidth{0pt} \tablecolumns{5}
\tablenum{2} \tablecaption{Data for clumps swA \& swB} \tablehead{
\colhead{Transition} & \colhead{v\scriptsize{$_{\textrm{lsr}}$}} & \colhead{$\Delta$v} & \colhead{S\scriptsize{$_{\textrm{peak}}$}} & \colhead{$\int$ S$_{\nu}$ dv} & \colhead{Model} \\
 & \colhead{\phn(km s$^{-1}$)\phn} & \colhead{(km s$^{-1}$)} & \colhead{\phn(Jy)\phn} & \colhead{(Jy km s$^{-1}$)} & \colhead{(Jy km s$^{-1}$)} }
\startdata \cutinhead{clump swA}
2$_0$--1$_0$ A$^+$  & $-$45.9 & 3.3 & 1.8 & 6.0$\pm$0.6 & 5.7 \\
2$_{-1}$--1$_{-1}$ E    & $-$45.9 & 3.6 & 1.4 & 5.1$\pm$0.5 & 4.9 \\
0$_0$--1$_{-1}$ E   & $-$46.7 & 4.6 & 0.4 & 1.6$\pm$0.3 & 0.8 \\
2$_0$--1$_0$ E  & $-$46.3 & 3.4 & 0.6 & 2.0$\pm$0.2 & 2.3 \\
2$_1$--1$_1$ E  & $-$46.3 & 3.1 & 0.3 & 1.0$\pm$0.2 & 0.4 \\
5$_{-1}$--4$_0$ E   & $-$45.8 & 2.1 & 2.2 & 5.3$\pm$0.5 & 5.0 \\
8$_0$--7$_1$ A$^+$  & $-$46.4 & 2.5 & 2.0 & 5.2$\pm$0.5 & 5.6 \\
\cutinhead{clump swB}
2$_0$--1$_0$ A$^+$  & $-$46.4 & 2.8 & 3.1 & 9.5$\pm$1.0 & 8.1 \\
2$_{-1}$--1$_{-1}$ E    & $-$46.6 & 2.5 & 2.5 & 6.7$\pm$0.7 & 6.5 \\
0$_0$--1$_{-1}$ E   & $-$46.6 & 2.4 & 1.4 & 3.6$\pm$0.4 & 3.2 \\
2$_0$--1$_0$ E  & $-$46.4 & 2.3 & 1.8 & 4.3$\pm$0.4 & 5.1 \\
2$_1$--1$_1$ E  & $-$46.4 & 2.3 & 1.0 & 2.4$\pm$0.3 & 2.1 \\
5$_{-1}$--4$_0$ E   & $-$46.4 & 2.2 & 4.0 & 9.4$\pm$1.0 & 4.5 \\
8$_0$--7$_1$ A$^+$  & $-$46.2 & 2.7 & 2.2 & 6.1$\pm$0.6 & 6.7 \\
\enddata
\end{deluxetable}

\clearpage


\begin{deluxetable}{lccccc}
\tablewidth{0pt} \tablecolumns{6} \tablenum{3}
\tablecaption{Physical parameters for clumps swA \& swB and
W3(H$_2$O)} \tablehead{
\colhead{Source} & \colhead{T\scriptsize{$_{\textrm{kin}}$}} & \colhead{n\scriptsize{$_{\textrm{H$_2$}}$}} & \colhead{N\scriptsize{$_{\textrm{A}}$}\normalsize+N\scriptsize{$_{\textrm{E}}$}} & \colhead{M\scriptsize{$_{\textrm{gas}}$}} & \colhead{(N\scriptsize{$_{\textrm{A}}$}\normalsize+N\scriptsize{$_{\textrm{E}}$}\normalsize)/N\scriptsize{$_{\textrm{H$_2$}}$}\normalsize} \\
& \colhead{(K)} & \colhead{(cm$^{-3}$)} & \colhead{(cm$^{-2}$)} &
\colhead{(M$_\sun$)} & } \startdata swA & 103$\pm$34    &
$\phn1.4^{+1.2}_{-0.6} \times 10^5\phn$   & $3.4^{+2.3}_{-1.4}
\times 10^{15}$ & $\approx0.05$ & $3.7^{+4.4}_{-2.0}\times10^{-7}$
\tablenotemark{a} \\ [4pt] swB & $\phn$47$\pm\phn$6    &
$2.0^{+3.3}_{-1.2} \times 10^6$   & $4.6^{+1.3}_{-1.0} \times
10^{15}$ & $\approx1.1\phn$ & $3.1^{+4.0}_{-1.7}\times10^{-8}$
\tablenotemark{a} \\ [4pt]
W3(H$_2$O) & 140$\pm$10 & $\phn0.9^{+2.4}_{-0.7} \times 10^7\phn$   & $2.4^{+0.3}_{-0.3} \times 10^{18}$ & $\approx1.6\phn$ & $5.\phn^{+18}_{-4}\phd\times10^{-6}$ \tablenotemark{a} \\
 & \nodata & \nodata & \nodata & \nodata & $1.7^{+1.3}_{-0.7} \times 10^{-6}$ \tablenotemark{b} \\ [2pt]
\enddata
\tablenotetext{a}{N\scriptsize{$_{\textrm{H$_2$}}$}\normalsize is
based on density and estimated path length.  The uncertainty in
the fractional abundance typically is dominated by the uncertainty
in density.} \tablenotetext{b}{Estimated fractional abundance
based on infrared pumping. The stated uncertainty reflects only
that associated with the amount of infrared pumping and ignores
uncertainties in the gas-to-dust ratio, the dust absorption
coefficient, and the dust opacity law.} \tablecomments{Estimated
68\% confidence limits.}
\end{deluxetable}

\clearpage


\begin{deluxetable}{lclcc}
\tabletypesize{\footnotesize} \tablewidth{300pt} \tablecolumns{5}
\tablenum{4} \tablecaption{Integrated Fluxes from W3(H$_2$O)}
\tablehead{
\colhead{Transition} & \colhead{Frequency} & \colhead{E$_u$} & \colhead{$\int$ S$_{\nu}$ dv (obs.)\tablenotemark{a}} & \colhead{ Model }\\
& \colhead{(MHz)} & \colhead{(K)} & \colhead{(Jy km s$^{-1}$)} &
\colhead{(Jy km s$^{-1}$)}} \startdata
7$_2$--8$_1$ A$^-$   & $\phn80993.26$   & 103. & $\phn3.5\pm\;\;$0.4 & $\phn3.7$ \\
13$_{-3}$--14$_{-2}$ E & $\phn84423.71$ & 274. & $\phn2.2\pm\;\;$0.2 & $\phn2.5$ \\
5$_{-1}$--4$_0$ E    & $\phn84521.21$   & $\phn40.$ & $\phn7.2\pm\;\;$0.7 & $\phn7.1$ \\
6$_{-2}$--7$_{-1}$ E & $\phn85568.07$   & $\phn75.$ & $\phn5.6\pm\;\;$0.6 & $\phn4.7$ \\
7$_2$--6$_3$ A$^-$   & $\phn86615.60$   & 103. & $\phn3.3\pm\;\;$0.3 & $\phn3.0$ \\
7$_2$--6$_3$ A$^+$   & $\phn86902.95$   & 103. & $\phn2.8\pm\;\;$0.3 & $\phn3.1$ \\
8$_{-4}$--9$_{-3}$ E & $\phn89505.78$   & 171. & $\phn3.1\pm\;\;$0.4 & $\phn3.1$ \\
1$_0$--2$_1$ E $\phn\phn$ v$_t$=1 & $\phn93196.66$ & 303. & $\phn1.8\pm\;\;$0.2 & $\phn1.3$ \\
8$_3$--9$_2$ E       & $\phn94541.81$   & 131. & $\phn4.8\pm\;\;$0.5 & $\phn5.0$ \\
8$_0$--7$_1$ A$^+$   & $\phn95169.52$   & $\phn84.$ & 11.5$\pm\;\;$1.2 & $\phn8.6$ \\
2$_1$--1$_1$ A$^+$ $\phd$ v$_t$=1 & $\phn96396.06$ & 332. & $\phn1.4\pm\;\;$0.2 & $\phn1.0$ \\
2$_1$--1$_1$ E $\phn\phn$ v$_t$=1 & $\phn96492.16$ & 298. & $\phn1.7\pm\;\;$0.2 & $\phn1.3$ \\
2$_0$--1$_0$ E $\phn\phn$ v$_t$=1 & $\phn96493.55$ & 308. & $\phn2.0\pm\;\;$0.2 & $\phn1.6$ \\
2$_{-1}$--1$_{-1}$ E  v$_t$=1 & $\phn96501.70$ & 420. & $\phn0.8\pm\;\;$0.2 & $\phn0.6$ \\
2$_0$--1$_0$ A$^+$ $\phd$ v$_t$=1 & $\phn96513.67$ & 431. & $\phn0.7\pm\;\;$0.2 & $\phn0.7$ \\
2$_1$--1$_1$ A$^-$ $\phd$ v$_t$=1 & $\phn96588.59$ & 332. & $\phn1.3\pm\;\;$0.2 & $\phn1.0$ \\
2$_{-1}$--1$_{-1}$ E & $\phn96739.36$   & $\phn12.5$ & $\phn6.8\pm\;\;$0.7 & $\phn6.4$ \\
2$_0$--1$_0$ A$^+$   & $\phn96741.38$   & $\phn\phn7.0$ & $\phn9.2\pm\;\;$0.9 & $\phn7.2$ \\
2$_0$--1$_0$ E       & $\phn96744.55$   &  $\phn20.$ & $\phn8.1\pm\;\;$0.8 & $\phn7.1$ \\
2$_1$--1$_1$ E       & $\phn96755.51$   &  $\phn28.$ & $\phn6.8\pm\;\;$0.7 & $\phn6.1$ \\
6$_1$--5$_0$ E $\phn\phn$ v$_t$=1 & $\phn99730.96$ & 340. & $\phn2.4\pm\;\;$0.4 & $\phn2.5$ \\
13$_2$--12$_3$ E     & 100638.87        & 234. & $\phn4.5\pm\;\;$0.5 & $\phn5.6$ \\
13$_{-3}$--12$_{-4}$ E & 104060.72      & 274. & $\phn2.3\pm\;\;$0.8 & $\phn3.7$ \\
11$_{-1}$--10$_{-2}$ E & 104300.40      & 159. & $\phn6.8\pm\;\;$0.7 & $\phn7.7$ \\
13$_{-2}$--13$_1$ E  & 104336.64        & 237. & $\phn3.4\pm\;\;$1.1 & $\phn2.4$ \\
10$_4$--11$_3$ A$^-$ & 104354.86        & 208. & $\phn3.7\pm\;\;$0.8 & $\phn4.4$ \\
10$_4$--11$_3$ A$^+$ & 104410.49        & 208. & $\phn3.4\pm\;\;$0.6 & $\phn4.4$ \\
13$_1$--12$_2$ A$^+$ & 105063.76        & 224. & $\phn7.1\pm\;\;$0.7 & $\phn7.3$ \\
14$_{-2}$--14$_1$ E  & 105576.39        & 270. & $\phn3.2\pm\;\;$0.3 & $\phn2.8$ \\
3$_1$--4$_0$ A$^+$   & 107013.81   & $\phn28.$ & $\phn7.5\pm\;\;$0.8 & $\phn9.9$ \\
15$_{-2}$--15$_1$ E  & 107159.92        & 305. & $\phn3.6\pm\;\;$0.4 & $\phn3.3$ \\
0$_0$--1$_{-1}$ E    & 108893.93        & $\phn13.1$ & $\phn6.9\pm\;\;$0.7 & $\phn7.3$ \\
14$_5$--15$_4$ E     & 109138.69        & 380. & $\phn4.3\pm\;\;$0.8 & $\phn2.9$ \\
16$_{-2}$--16$_1$ E  & 109153.21        & 342. & $\phn3.6\pm\;\;$0.4 & $\phn3.5$ \\
7$_2$--8$_1$ A$^+$   & 111289.60        & 103. & $\phn8.1\pm\;\;$0.8 & $\phn8.1$ \\
5$_1$--4$_2$ E       & 216945.56        & $\phn56.$ & 28.0$\pm\;\;$7.0 & 37.8\\
6$_1$--7$_2$ A$^-$ $\phd$ v$_t$=1 & 217299.16  & 374. & 15.5$\pm\;\;$3.9 & 24.2\\
8$_0$--7$_1$ E       & 220078.49        & $\phn97.$ & 41.4$\pm$10.4 & 47.6 \\
\enddata
\tablenotetext{a}{Integrated over the velocity range from $-$54 to
$-44$ \kmps.}

\end{deluxetable}

\clearpage

\figcaption{
Emission from lines with upper state energies less than 100 K for which
data are available down to spatial frequencies of 3 k$\lambda$ or less.
Lines are presented in order of increasing upper state energy.
The map center is $\alpha$(J2000) = 2$^\mathrm{h}$27$^\mathrm{m}$03\fs87,
$\delta$(J2000) = 61$\arcdeg$52$'$24\farcs6,
the nominal position of W3(OH).
\cite{S01} found the 3 mm continuum from the W3(OH) \UCHII\ region
to peak 0\farcs2 $\pm$ 0\farcs1 west of this position.
The TW object is 5\farcs94 east of the phase center \citep{W99,R95}.
With the exception of panel a, the plots are of average intensity
over a \vlsr\ range of $-$53 to $-$42 \kmps.
Panel a is a blend of lines, and the intensity is an average
over a range of channels corresponding to $-$52.3 \kmps\ for the
2$_0$--1$_0$ A$^+$ line to $-$41.6 \kmps\ for the
2$_{-1}$--1$_{-1}$ E line.
Beams are shown in the lower left corners of the plots,
and their sizes are listed in Table 1.
Contours are .05, .1, .15, .2, .25, .3, .4, .5, .6 Jy beam$^{-1}$
for panels a, c, d, e, and f
and .075, .15, .225, .3, .4, .5, .6 Jy beam$^{-1}$ for panel b.
}

\figcaption{
J = 2--1 lines from the first torsionally excited state.
Map center and velocity range are the same as for Figure 1.
Beams are shown in the lower left corners of the plots,
and their sizes are listed in Table 1.
Symbols mark the positions of the TW object and
the 3 mm continuum center of W3(OH).
Contours are multiples of 0.05 Jy beam$^{-1}$.
}

\figcaption{
Maps of 15 b-type lines towards W3(OH)/W3(H$_2$O).
Map center and velocity range are the same as for Figure 1.
Beams are shown in the lower left corners of the plots,
and their sizes are listed in Table 1.
Where space allows, symbols mark the positions of the TW object
and the 3 mm continuum center of W3(OH).
Contours are .05, .1, .15, ..., .4, .5, .6, .7, .8 Jy beam$^{-1}$
for panels e, f, g, i, k, l, m, and n.
Contours are .125, .25, .375, .5, .75, 1., 1.5, 2.
and then multiples of 1. Jy beam$^{-1}$ for panels a, c, h, and j.
For panels b, d, and o, the contours are multiples of 0.25 Jy beam$^{-1}$.
}

\figcaption{
Continuum map at 1.3 mm with 0.6 arcsec resolution.
Contours are multiples of 0.1 Jy beam$^{-1}$.
Peaks A, B, and C of \cite{W99} are labelled.
Our measured position and positional uncertainty of the 3 mm
W3(OH) continuum center is indicated by the filled box.
The $2.0\times1.2$ arcsec ellipse represents our best estimate
of the position and extent of the region of methanol emission
near W3(H$_2$O).
}

\figcaption{
Rotation diagram fit to methanol emission from W3(H$_2$O).
In panel (a) the ordinate is
$3c^2 \int S_{\nu}dv \;/\;
(16\pi^3\nu^3 \Omega \mu^2S)$.
In panel (b) the ordinate is the above quantity times C$_{\tau}$,
that is, corrected for optical depth.
The solid line is the best weighted $\chi^2$ fit to the data,
and gives a rotational temperature of 149 K
and N$_0$/g$_0$ = $1.3 \times 10^{15}$ \cmsq.
In panel (c) the ordinate is the ratio of the observed integrated flux
to that predicted by the non-LTE model discussed in the text.
}

\figcaption{
Spectra of three lines from a 4 x 4 arcsec region around the
center of W3(H$_2$O), offset by ($\Delta \alpha$ = +5.3,
$\Delta \delta$ = +0.1) arcsec from the
reference position.
}

\newpage
\plotone{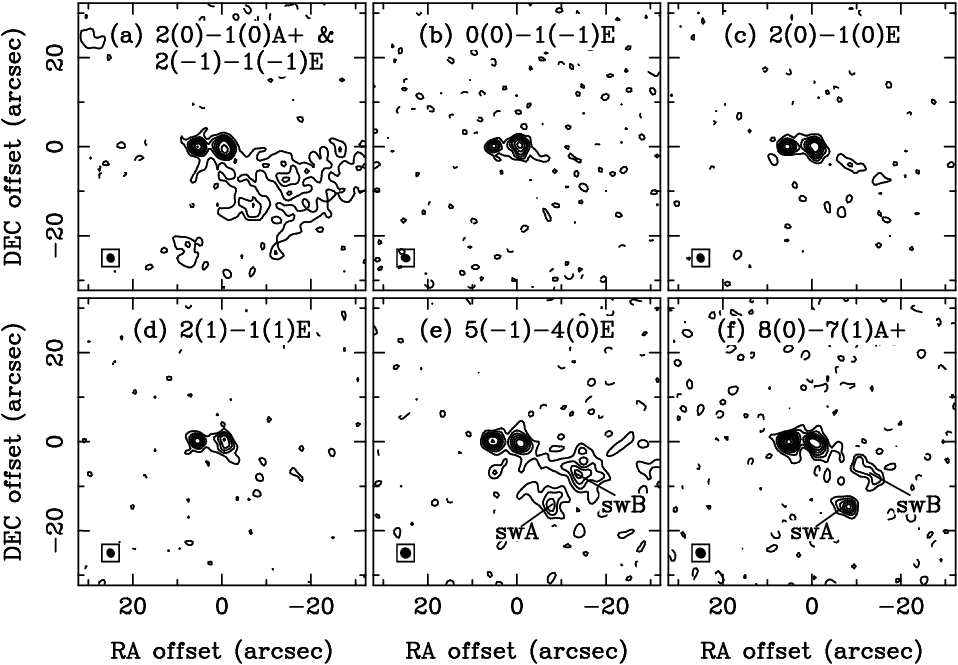}
\newpage
\plotone{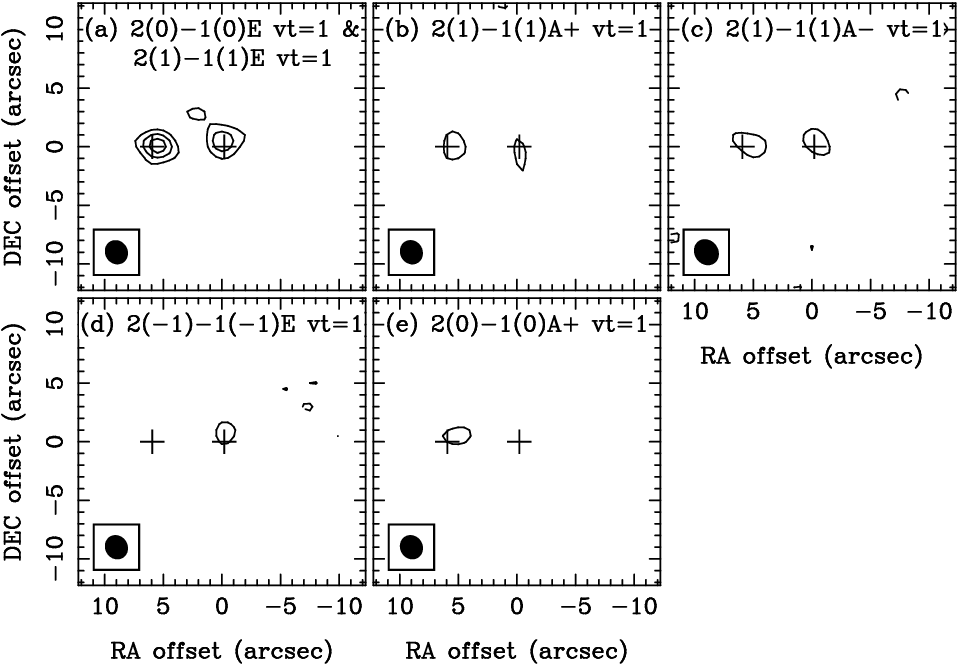}
\newpage
\epsscale{.80}
\plotone{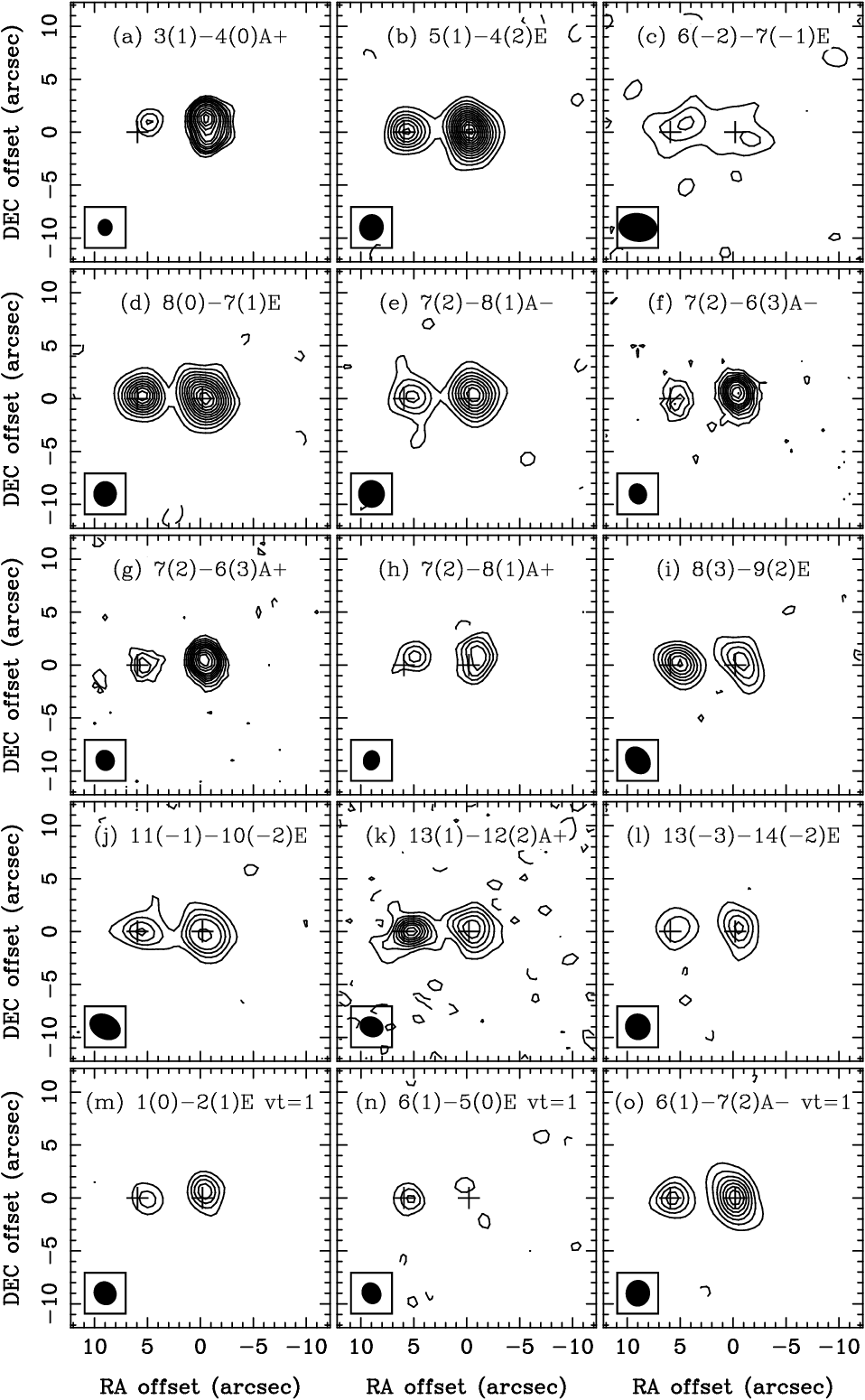}
\newpage
\epsscale{1}
\plotone{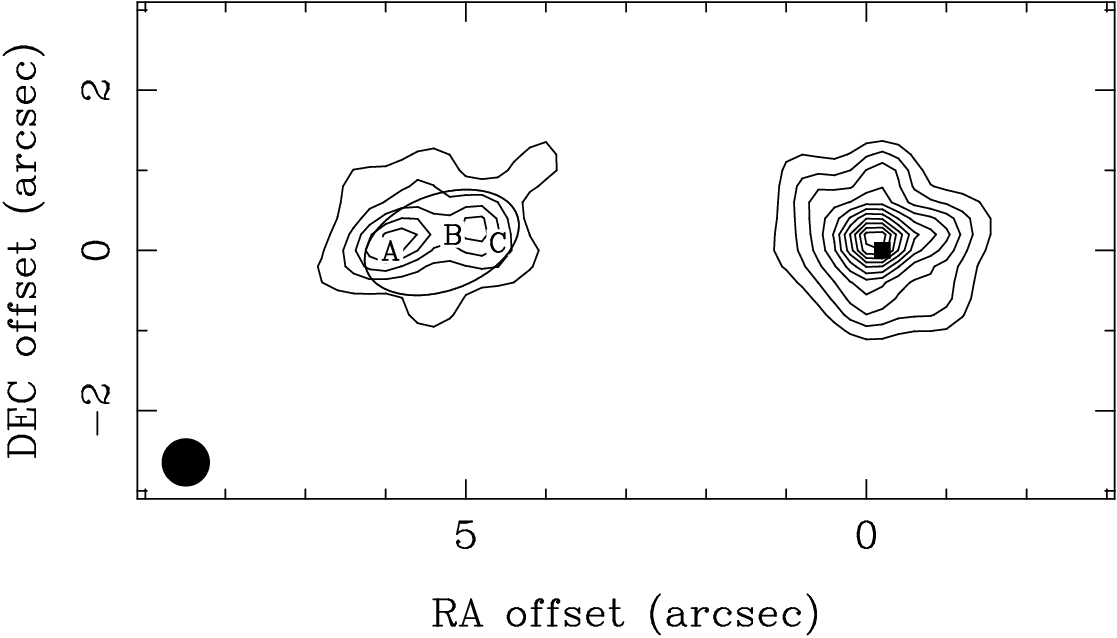}
\newpage
\plotone{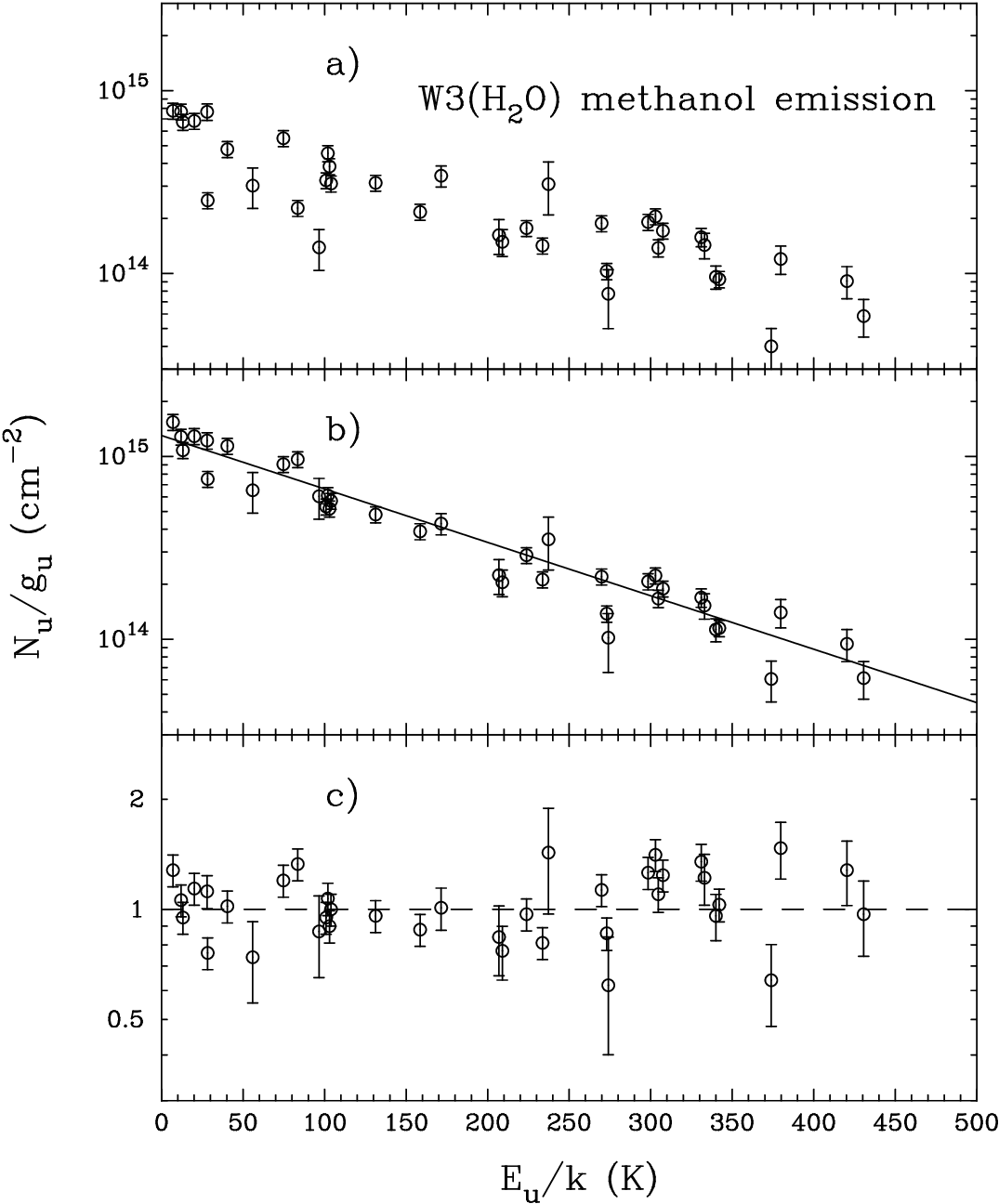}
\newpage
\plotone{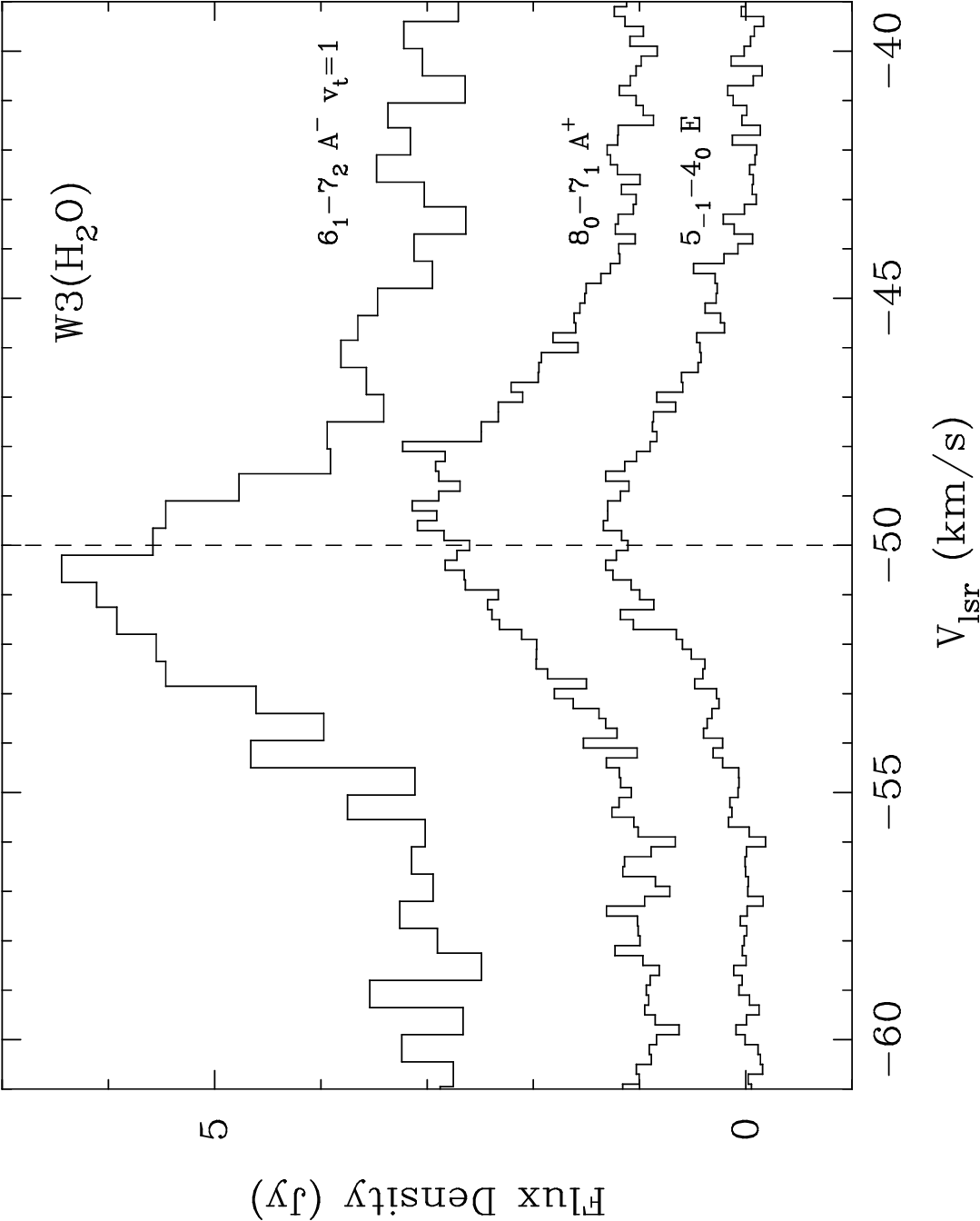}

\end{document}